\documentclass{article}
\usepackage{spconf,amsmath,graphicx}

\usepackage[T1]{fontenc} 
\usepackage{amsmath,amssymb,amsthm}
\usepackage[cmintegrals]{newtxmath}
\usepackage{bm} 
\usepackage{float}
\usepackage{cite}
\usepackage{graphicx,subfigure,epstopdf}
\usepackage{multirow}
\usepackage{url}
\usepackage{algorithm}
\usepackage{algpseudocode,color,xcolor}

\newtheorem{thm}{Theorem}

\newcommand{\R}{\mathbb{R}}
\newcommand{\C}{\mathbb{C}}

\newcommand{\vct}[1]{\boldsymbol{#1}}
\newcommand{\mtx}[1]{\boldsymbol{#1}}

\newcommand{\<}{\langle}
\renewcommand{\>}{\rangle}

\newcommand{\set}[1]{\mathcal{#1}}

\newcommand{\ve}{\vct{e}}
\newcommand{\vf}{\vct{f}}
\newcommand{\vg}{\vct{g}}
\newcommand{\vh}{\vct{h}}

\newcommand{\vm}{\vct{m}}

\newcommand{\vr}{\vct{r}}
\newcommand{\vs}{\vct{s}}

\newcommand{\vu}{\vct{u}}
\newcommand{\vv}{\vct{v}}

\newcommand{\vx}{\vct{x}}
\newcommand{\vy}{\vct{y}}
\newcommand{\vz}{\vct{z}}
\newcommand{\mF}{\mtx{F}}

\newcommand{\mI}{\mtx{I}}

\newcommand{\mR}{\mtx{R}}

\newcommand{\setA}{\set{A}}

\newcommand{\setN}{\set{N}}
\newcommand{\setO}{\set{O}}

\newcommand{\setT}{\set{T}}

\def\tf{\tilde{F}}

\allowdisplaybreaks

\title{Channel Protection using Random Modulation}
\name{Ali Ahmed\thanks{A.\ A.\ is a visiting Assistant Professor at the Department of Mathematics at MIT in Cambridge, MA, and Assistant Professor at the Department of Electrical Engineering, Information Technology University (ITU), Lahore. Email: \texttt{alikhan@mit.edu.} This work was supported by the Higher Education Commission (HEC), Pakistan under the National Research Program for Universities (NRPU), Project no. 6856. Manuscript submitted on October 30, 2017.}, Humera Hameed}
\address{Department of Electrical Engineering,\\
	Information Technology University,Lahore, Pakistan\\
	\{ali.ahmed,phdee17001\}@itu.edu.pk}
\begin{document}
\ninept
\maketitle
\begin{abstract}
    This paper shows that modulation protects a bandlimited signal against convolutive interference.  A signal $s(t)$, bandlimited to $B$Hz, is modulated (pointwise multiplied) with a known random sign sequence $r(t)$, alternating at a rate $Q$, and  the resultant \textit{spread spectrum} signal $s(t) \odot r(t)$ is convolved against an $M$-tap channel impulse response $h(t)$ to yield the observed signal $y(t)= (s(t)\odot r(t))\circledast h(t),$ where $\odot$ and $\circledast$ denote pointwise multiplication, and circular convolution, respectively. 
    
    We show that both $s(t)$, and $h(t)$ can be provably recovered using a simple gradient descent scheme by alternating the binary waveform $r(t)$ at a rate $Q \gtrsim B + M$(to within log factors and a signal coherences) and sampling $y(t)$ at a rate $Q$. We also present a comprehensive set of phase transitions to depict the trade-off between $Q$, $M$, and $B$ for successful recovery. Moreover, we show stable recovery results under noise. 
\end{abstract}

\section{Introduction} \label{sec:introduction}
This paper shows that a simple modulation of a bandlimited signal protects it against unknown convolutive channel interference. We take a periodic signal $s(t)$ in $t \in [0,1)$, bandlimited to $B$ Hertz. We can express $s(t)$ in Fourier basis as
\begin{align}\label{eq:bandlimited-signal}
s(t)=\sum_{k=-B}^{B} x[k] e^{\iota 2\pi kt}, \ t \in [0,1)
\end{align}
Shannon-Nyquist sampling theorem tell us that the signal $s(t)$ can be captured by uniformly sampling at a rate $Q \geq K = 2B+1$. We modulate $s(t)$ by pointwise multiplying it with a known random binary waveform $r(t)$, alternating at a rate $Q$, and use $r(t)\odot s(t)$ to denote the modulated signal, where $\odot$ represents the pointwise multiplication. The modulated signal is then filtered via an unknown LTI system characterized by an impulse response $h(t)$ resulting in the convolution $y(t) =  (r(t) \odot s(t))\circledast h(t)$, where $\circledast$ denotes the circular convolution. We observe $y(t)$, and aim to recover both $s(t)$, and $h(t)$. Our main result shows that $s(t)$, and $h(t)$ can be recovered using a simple gradient descent algorithm. Moreover, we are able to quantify exactly the alternating rate $Q$ of the binary waveform, and the sampling rate, which is also fixed at $Q$, of $y(t)$ sufficient to achieve the exact recovery of $s(t)$, and $h(t)$. 

Modulation is prevalently used in signal processing and communications to, for example, effectively use available spectral bands by multiplexing signals in different frequency bands \cite{goldsmith2005wireless}. Perhaps more closely related to our framework is code division multiple access \cite{Buehrer06CDMA}, where also a binary waveform is pointwise multiplied with data signal primarily to mitigate additive interferences from other user data signals.  In this paper, however, we are claiming that binary modulation also protects a signal against more involved and unknown convolutive interference. In other words, a premodulation step enables us to deconvolve the signal, and  impulse response without knowing either of them. In general, deconvolution of two unknown inputs is referred to as blind deconvolution (BD) and is severely ill-posed. BD is one of the fundamental problems in signal processing, communications, system theory, etc, \cite{moulines1995subspace,campisi2016blind} and this paper shows that a simple modulation step, which already happens to be a part of the signal preprocessing assembly line prior to transmission in many wireless communication systems, enables us to undo the convolution with an unknown impulse response at the receiver. 

\subsection{Discrete Form}

The bandlimited signal $s(t)$ in \eqref{eq:bandlimited-signal} can be captured taking $Q \geq K = 2B+1$ equally spaced samples at time instants $\setT_Q := \{\tfrac{q-1}{Q}, \ q\in [Q]\}$, where, in general, $[Q] = \{1,2,3,\ldots, Q\}$.  Let $\mF$ be a $Q \times Q$ DFT matrix with entries $F[q,k] = \frac{1}{\sqrt{Q}} \mathrm{e}^{-j 2\pi kq/Q}, \ (q,k) \in [Q]\times [Q]$. In general, we will denote by$\mF_J$, a submatrix formed by selecting the first $J$ columns of $\mF$. Similarly, the notation $\mF^*_J$ denotes a submatrix consisting of the first $J$ columns of $\mF^*$ where $\mF^*$ denotes hermitian of $\mF$. Then the samples of $s(t)$ can be expressed as 
\[
\vs=\mF^*_{K}\vx,
\]
where the entries of the $K$-vector $\vx$ are the Fourier coefficients. Let $\vr$ be a $Q$-vector of samples of binary waveform $r(t)$ in $\setT_Q$. We model the impulse response $h(t)$ as an $M$-tap filter 
\begin{align}\label{eq:impulse-response}
h(t)= \sum_{m=1}^M h[m] \delta (t-t_m),
\end{align}
where each $t_m \in \setT_Q$ is unique and known. Assuming $h[m]$ to be the $m$th entry of an $M$-vector $\vh$. The rate $Q$ samples of $y(t)$ on the grid $t \in \setT_Q$ are collected in $Q$-vector $\vy$  below 
\begin{align}\label{eq:model}
\vy = (\vr \odot \vs) \circledast \vh,
\end{align}
where $\circledast$ denotes the $Q$-point circular convolution\footnote{The $Q$-point circular convolution $\vx_1 \circledast \vx_2$ of $\vx_1 \in \C^Q$, and $\vx_2 \in \C^M$, where $M \leq Q$, is the circular convolution of $Q$-vector $\vx_1$, and $\vx_2$ zero padded to length $Q$. Mathematically, $\vx_1 \circledast \vx_2 = \mF^* \text{diag}(\hat{\vx}_1) \mF_M \vx_2$, where $\hat{\vx}_1 = \sqrt{Q} \mF\vx_1$.}, and $\vy \in \mathbb{C}^Q$. The objective is to recover the unknown $\vx$, and $\vh$ from $\vy$. 

Formally, we take the measurements in the Fourier domain 
\begin{align}\label{eq:measurements}
\hat{\vy} &= \mF\vy= \sqrt{Q}\big( \mF(\vr\odot \vs_{0})\odot \mF_M\vh_0\big) + \ve\notag\\
&= \sqrt{Q}( \mF_Q\mR\mF^*_K\vx_{0}\odot\mF_M\vh_0) + \ve,
\end{align}
where $\mR := \text{diag}(\vr)$ is a $Q\times Q$ diagonal matrix, $\vh_0$, $\vx_0$ are the ground truths, and $\ve \in \C^Q$ denote the additive noise in the Fourier domain. To deconvolve, we minimize the measurement loss by taking a gradient step in each of the unknowns $\vh_0$, and $\vx_{0}$ while keeping the other fixed. This paper details a particular set of conditions on the sample complexity, subspace dimensions, and the signals/filters under which this computationally feasible alternating gradient descent scheme provably succeeds.

Notice that the measurements in \eqref{eq:measurements} are non-linear in the unknowns $(\vh_0,\vx_0)$, however, are linear in the rank-1 outer-product $\vh_0\bar{\vx}_0^*$ where $\bar{\vx}_0^*$ denotes conjugate hermitian of $\vx_0$.  To see this, let $\vf^*_q\in \C^M$ be the $q$th row of $\mF_M$ and $\hat{\vg}_q\in \C^{K}$ be the $q$th row of the  $Q \times K$ matrix $\sqrt{Q}(\mF_Q\mR\mF_K^*)$. The $q$th entry $\hat{y}[q]$ of measurements $\hat{\vy}$ in \eqref{eq:measurements} is then simply 
\begin{align}\label{eq:entrywise-measurements}
\hat{y}[q] = \vf_q^*\vh_0\bar{\vx}_0^*\hat{\vg}_q+e[q] = \< \vf_q\hat{\vg}_q^*,\vh_0\bar{\vx}_0^*\>+e[q];
\end{align}
the linearity of the measurements in $\vh_0\bar{\vx}_0^*$ is clear from the last equality above. We also define a linear map $\setA: \C^{M \times K} \rightarrow \C^{Q}$ that maps $\vh_0\vx_0^*$ to the vector $\hat{\vy}$. The action of $\setA$ on a rank-1 matrix $\vh\vx^*$ returns 
\begin{align}\label{eq:linear-map}
&\setA(\vh\vx^*) := \{\vf_q^*\vh \bar{\vx}^*\hat{\vg}_{q}\}_{q}, \ q  \in [Q], \notag\\
&\text{and therefore,} \ \hat{\vy} = \setA(\vh_0\vx_0^*) + \ve,
\end{align}
where we used the definition of $\setA$ to compactly express \eqref{eq:entrywise-measurements}.  


\subsection{Implementation Potential}
\begin{figure}[htb]
\begin{minipage}[b]{1.0\linewidth}
  \centering
  \centerline{\includegraphics[scale = 0.5, trim = -16cm 8cm 0cm 0.5cm,clip]{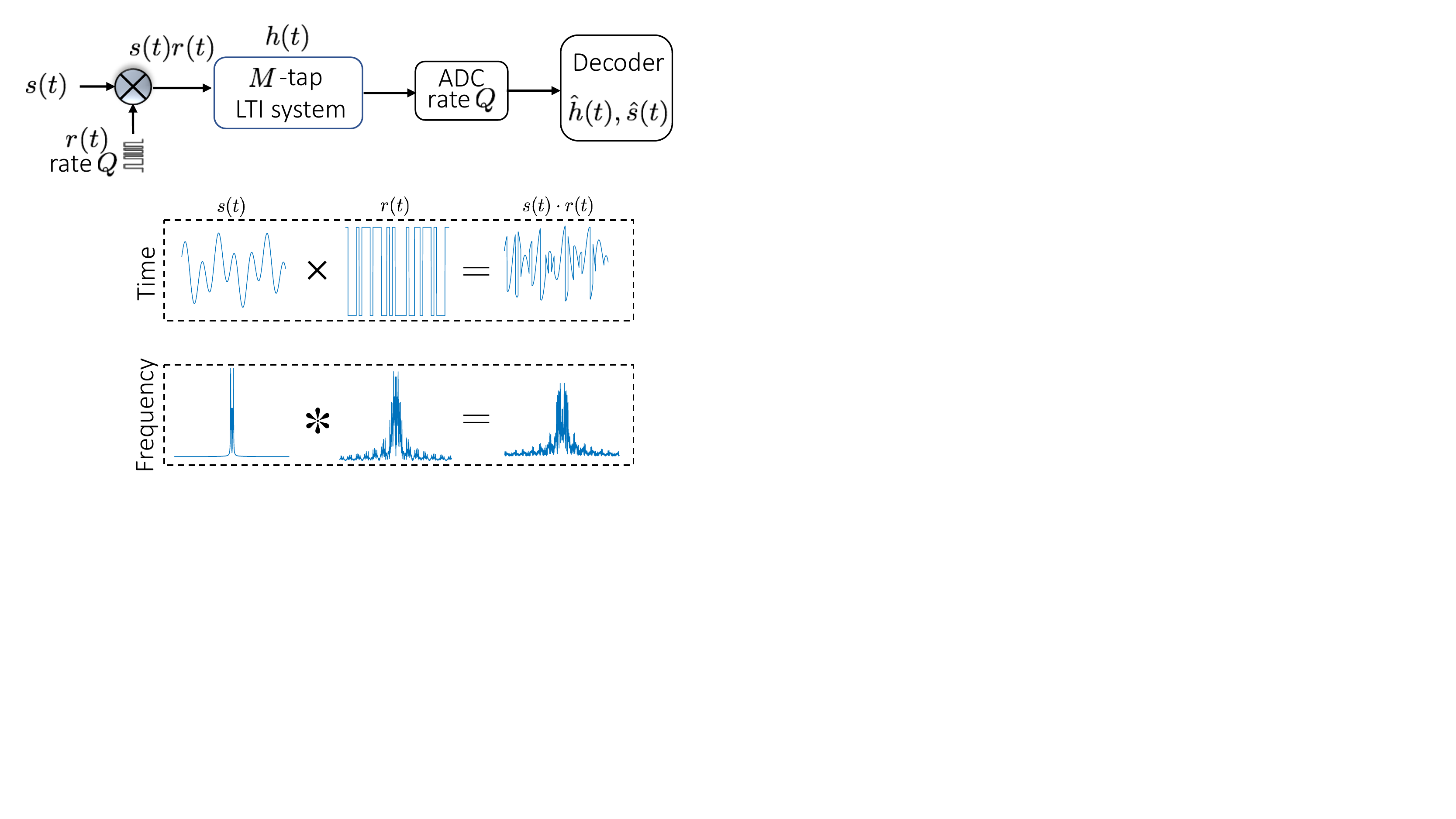}}
\end{minipage}
\caption{ \small\sl Analog implementation for real time protection against channel intereference.  A continuosus time signal $s(t)$, bandlimited to $B$ Hz, is modulated with a random binary waveform $r(t)$ alternating at a rate $Q$. The modulated signal drives an unknown LTI system characterized by an $M$-tap impulse response $h(t)$. The resulting signal is sampled at a rate $Q$. Operate the modulator and ADC at a rate $Q \gtrsim \max(B,M)$ (to within a constant, log factors and coherences), and recover $s(t)$, and $h(t)$ using gradient descent. Underneath, the preprocessing is shown in time, and frequency domain. Modulation in time domain spreads the spectrum, and the resulting higher frequency signal remains protected against the distortions caused by an unknown LTI system. }
\label{fig:Modulators}
\end{figure}

Binary modulation of an analog signal can be easily implemented using switches that flip the signs of the signal in real time; the setup is shown in Figure \ref{fig:Modulators}.

Fast rate binary switches can be easily implemented; see, for example, \cite{laska2007theory}, and  \cite{tropp2010beyond,ahmed2015compressive,ahmed2018compressive} for the use of binary switches in other applications in signal processing. The implementation potential combined with the ubiquity of blind deconvolution make this result interesting in system theory, applied communications, and signal processing, among other.

\subsection{Recent Literature}
  It is important to point out critical differences in the structural assumptions compared to the contemporary recent literature \cite{ahmed2015convex}, and \cite{ahmed2012blind,li2016rapid,lee2017spectral,ahmed2016leveraging,ahmed2018convex} on blind deconvolution, where the convolved signals are assumed to live in known subspaces spanned by the columns of a random Gaussian matrices.  Such random subspaces do not naturally arise in applications. We relinquish such restrictive Gaussian subspace assumption, and give a provable blind deconvolution result by only assuming random/generic sign (modulated) signal $\vs = \mF_K^*\vx$ that resides in realistic subspace spanned by the columns of  the DFT matrix. 
 


\subsection{Coherence Parameters}
Our main theoretical results depend on some signal dispersion measures that characterize how diffuse signals are in the Fourier domain. Intuitively, concentrated (not diffuse) signals in the Fourier domain in \textit{annihilate}  the measurements in \eqref{eq:measurements} making it relatively difficult (more stringent sample complexity requirements) to recover such signals. We refer to the signal diffusion measures as coherence parameters, defined and discussed below. 

For arbitrary vectors $\vh \in \C^M$, $\vx \in \C^{K}$, we define coherences 
\begin{align}\label{eq:muh-nux}
\mu_h^2 : = Q \frac{\|\mF_M\vh\|_\infty^2}{\|\vh\|_2^2}, \  \nu_x^2 := Q \frac{\|\mF_K^*\vx\|_\infty^2 }{\|\vx\|_2^2}.
\end{align}

Similar coherence parameters appear in the related recent literature on blind deconvolution \cite{ahmed2012blind,ahmed2016leveraging}, and elsewhere in compressed sensing \cite{candes09ex,candes2007sparsity}, in general. Without loss of generality, we assume that $\|\vh_0\|_2 = \sqrt{d_0}$, and $\|\vx_0\|_2 = \sqrt{d_0}$. For brevity, we will denote the coherence parameters $\mu^2_{h_0}$, and $\nu^2_{x_0}$ of the fixed ground truth vectors $(\vh_0,\vx_0)$ by
\begin{align}\label{eq:mu-nu}
\mu^2:=\mu^2_{h_0}, \ \text{and} \ \nu^2 := \nu^2_{x_0}. 
\end{align}

In words, coherence parameter $\mu_h^2$ is the peak value of the frequency spectrum of a fixed norm vector $\vh$. A higher value roughly indicates a concentrated spectrum and vice versa. It is easy to check that $1 \leq \mu_h^2 \leq Q$. 

On the other hand, $\nu_x^2$ quantifies the dispersion (not in the Fourier domain) of the signal $\vs = \mF_K^*\vx$. A signal concentrated in time (mostly zero) remains somewhat oblivious to the random sign flips $\vr \odot \vs$, and as a result is not as well-dispersed in the frequency domain. Let $\vg_{q}^*$ be the rows of $\mF_K^*$. By definition, $\nu_x^2\|\vx\|_2^2 \geq Q |\vg_{q}^*\vx|^2$ for any $q \in [Q]$. Summing over $q \in [Q]$ on both sides, and using the isometry of $\mF_K^*$ gives us the inequality $\nu_x^2 \geq 1$. The upper bound $\nu_x^2 \leq Q$ is easy to see using Cauchy Schwartz inequality, hence, $ 1 \leq \nu_x^2 \leq Q$. 


\begin{figure*}[htb]
	\centering
	\vspace{-0.5in}
	\begin{tabular}{ccc}
		\includegraphics[scale = 0.29, trim = 0cm 7.5cm 3cm 5cm,clip]{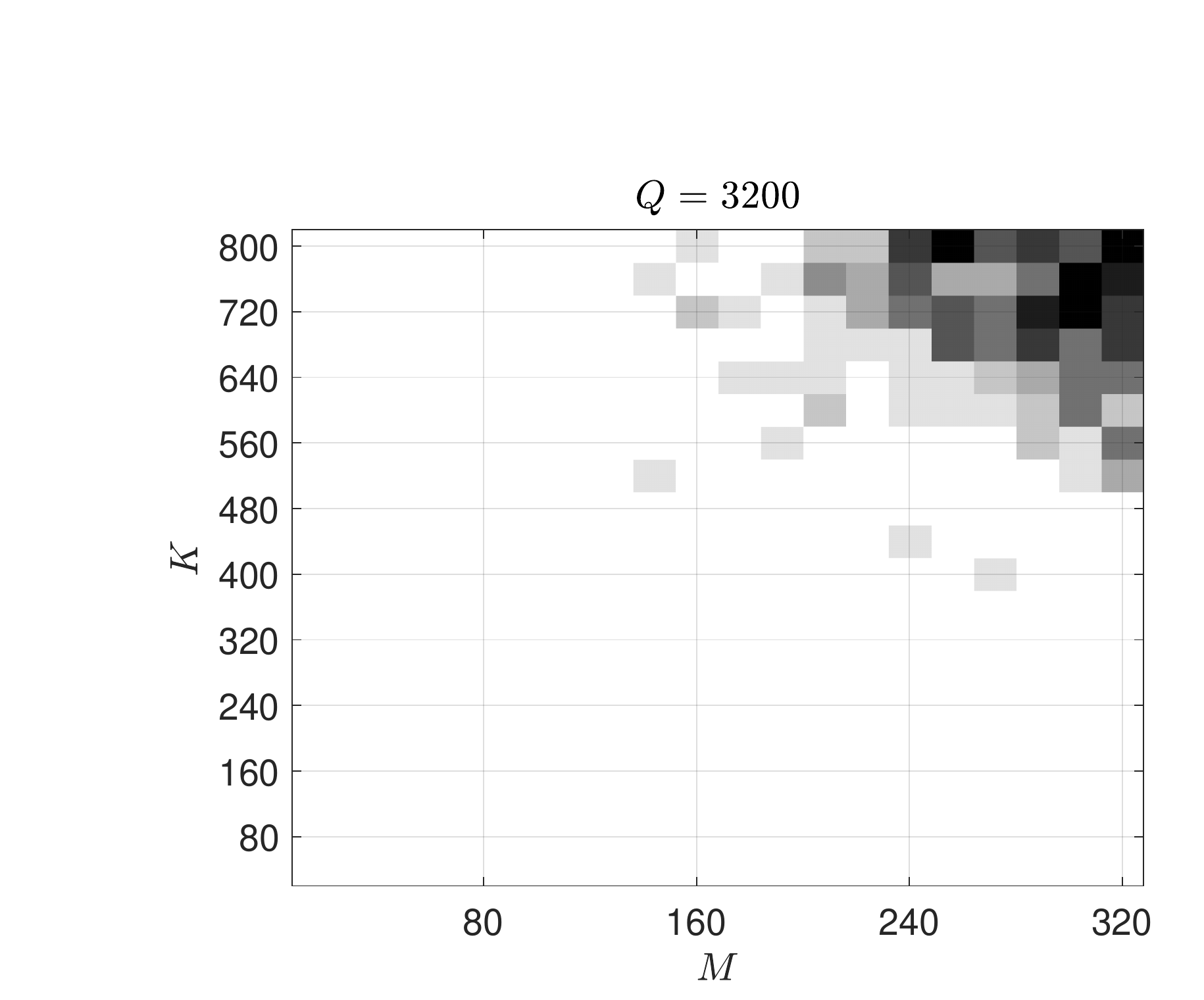}
		&\includegraphics[scale = 0.29, trim = 0cm 7.5cm 3cm 5cm,clip]{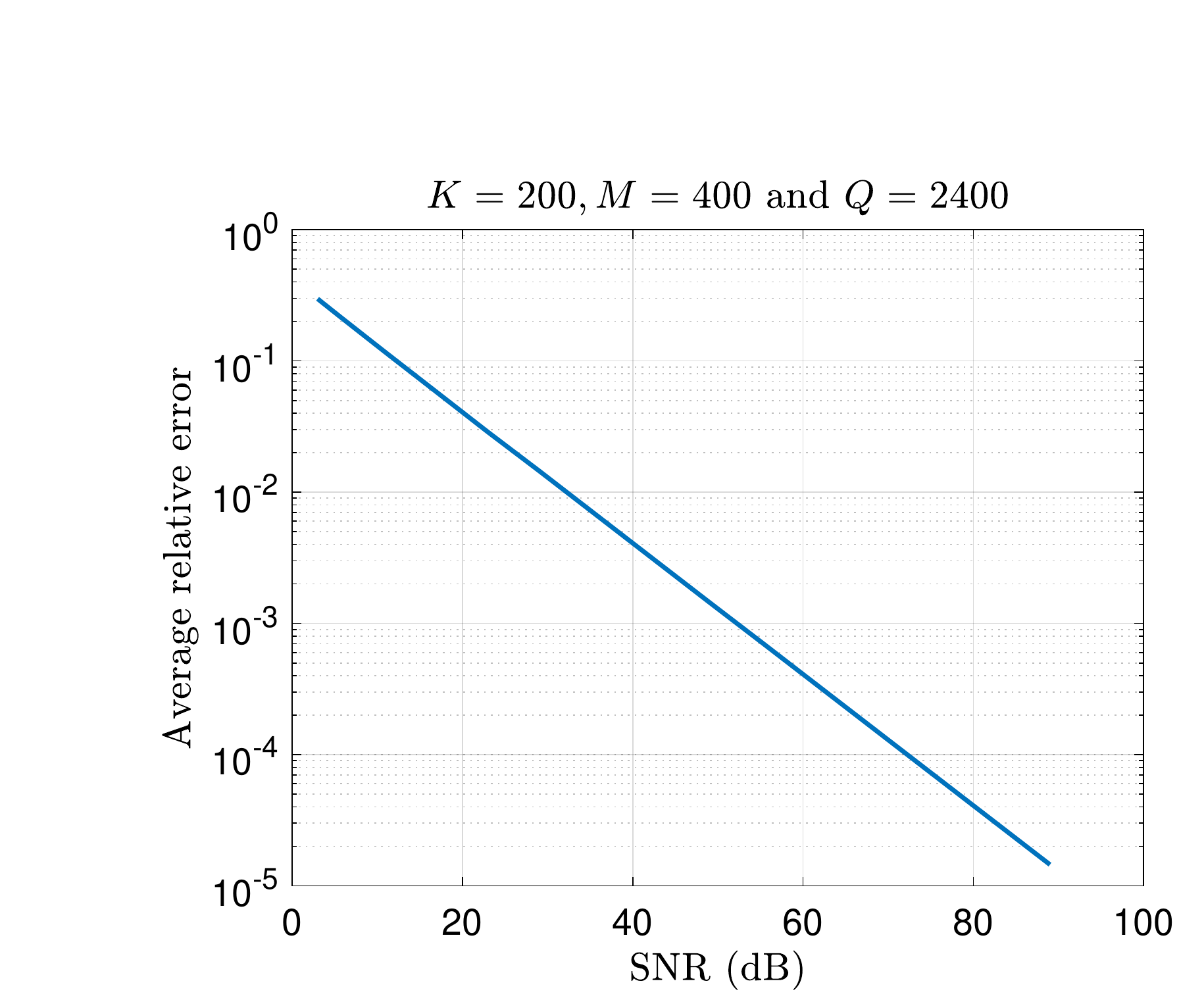}
		&\includegraphics[scale = 0.29, trim = 0cm 7.5cm -2cm 5cm,clip]{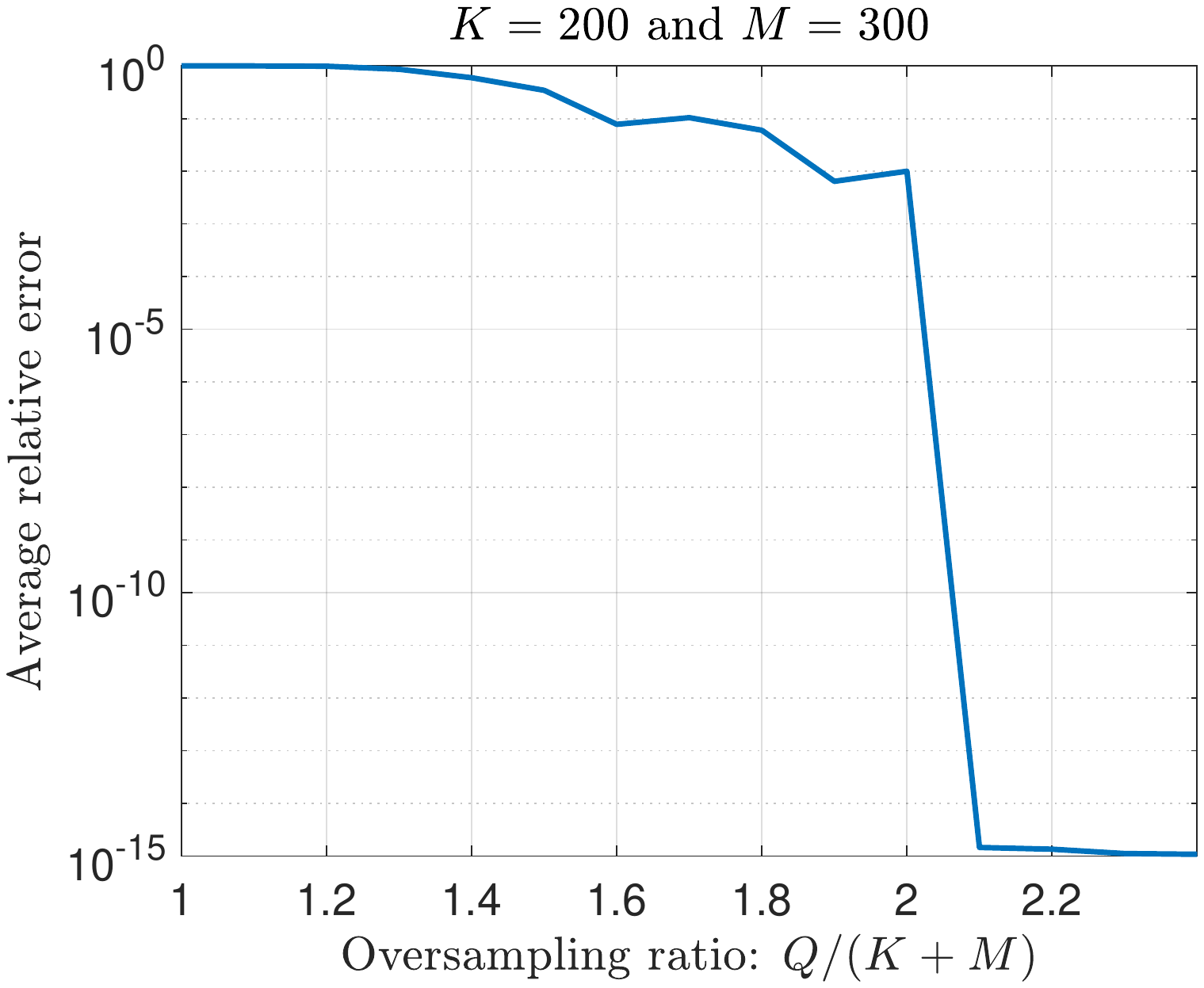}\\
	  \hspace{0.4in}	(a) &\hspace{0.4in} (b) & (c)\vspace{-0.1in}
	\end{tabular}
	\caption{\small \sl (a) Phase transition of $K$ vs. $M$ for fixed $Q$ show that for modulated inputs allow recovery with in log factor sum of $K$, and $M$, (b) The performance in the presence of additive meaurement noise, and (c) The number of samples  vs. the relative error. }
	\label{fig:Numerics}
\end{figure*}
\section{Recovery via Gradient Descent} \label{sec:Gradient-Descent}
Given measurements $\hat{\vy}$ of the ground truth $(\vh_0,\vx_0)$, we employ a regularized gradient descent algorithm that aims to minimize a loss function:
\begin{align}\label{eq:tF-def}
\tf(\vh,\vx) := F(\vh,\vx)+G(\vh,\vx).
\end{align}
w.r.t. $\vh$, and $\vx$, where the functions $F(\vh,\vx)$, and $G(\vh,\vx)$  account for the measurement loss, and regularization, respectively; and are defined below 
\begin{align}\label{eq:F-def}
& F(\vh,\vx) := \|\setA(\vh\vx^*) - \hat{\vy}\|_2^2 =  \|\setA(\vh\vx^*-\vh_0\vx_0^*)-\ve\|_2^2= \notag \\
&\|\setA(\vh\vx^*-\vh_0\vx_0^*)\|_2^2 + \|\ve\|_2^2 - 2 \text{Re}(\<\setA^*(\ve),\vh\vx^*-\vh_0\vx_0^*\>),
\end{align}
and
\begin{align}\label{eq:G-def}
& G(\vh,\vx) := \rho\Bigg[ G_0\left( \frac{\|\vh\|_2^2}{2d} \right)+ G_0\left(\frac{\|\vx\|_2^2}{2d} \right)+\sum_{q=1}^Q G_0\left( \frac{Q |\vf_q^* \vh|^2}{8d \mu^2}\right) \notag\\
&\qquad\qquad + \sum_{q=1}^Q G_0\left( \frac{Q |\vg_{q}^* \vx|^2}{8d \nu^2}\right)\Bigg], 
\end{align}
where $G_0 (z ) = \max\{z-1,0\}^2$. We set $\rho \geq d^2+\|\ve\|_2^2$, and $ 0.9 d_0 \leq d \leq 1.1 d_0$. Evidently, the regularizer $G(\vh,\vx)$ restricts the coherences $\mu_h^2$, $\nu_x^2$; and norms of $(\vh,\vx)$ under (to with in a constant) those of the ground truth $(\vh_0,\vx_0)$.

The proposed alternating gradient descent algorithm takes alternate Wirtinger gradient (of the loss function $\tf{(\vh,\vx)}$) steps in each of  the unknowns $\vh$, and $\vx$ while fixing the other; see Algorithm \ref{algo:gradient-descent} below for the pseudo code. The Wirtinger gradients are defined as\footnote{For a complex function $f(\vz)$, where $\vz = \vu + \iota \vv \in \C^L$, and $\vu, \vv \in \R^L$, the Wirtinger gradient is defined as $
\frac{\partial f}{\partial \bar{\vz}} = \frac{1}{2}\left( \frac{\partial f}{\partial \vu} + \iota \frac{\partial f}{\partial \vv}  \right).$} 
\begin{align}\label{eq:gFh-gFx-def}
\nabla \tilde{F}_{\vh} := \frac{\partial \tilde{F}}{\partial \bar{\vh}} = \frac{\overline{\partial \tilde{F}}}{\partial {\vh}}, \ \text{and} \ \nabla \tilde{F}_{\vx} := \frac{\partial \tilde{F}}{\partial \bar{\vx}} = \frac{\overline{\partial \tilde{F}}}{\partial {\vx}}.
\end{align}

Similar algorithms with provable recovery results appeared earlier beginning with \cite{jain2013low}, and then \cite{candes2015phase} for phase retrieval, and in \cite{li2016rapid} for blind deconvolution, however, with observation model different from \eqref{eq:model} considered here. For example, \cite{li2016rapid} assumes that $\vs$ lives in the span of the columns of a tall Gaussian matrix. We on the other hand do not make such a restrictive structural assumption, which is unfortunately rarely met in practice, and only assume that $\vs$ lives in a realistic subspace, spanned by the columns of the DFT matrix $\mF_K^*$ in this case. In addition, we comparatively assume much limited and structured randomness in the form of random signs of one of the convolved inputs. This random sign assumption might either be naturally satisfied or can be enforced in real time through an easy to implement signal modulation. This easy to realize modulation plus the deterministic subspace model broadens the applicability of our result in actual physical applications; for more details, see \cite{Ahmed2018Modulation}. Limited randomness, however, makes the analysis and proof of our theoretical results significantly more challenging than \cite{li2016rapid}. We have to resort to advanced techniques such as generic chaining \cite{talagrand2005generic} to control the some of the resultant random processes that arise in the proofs \cite{Ahmed2018Modulation}.

\begin{algorithm}[H]
	\caption{Wirtinger gradient descent with a step size $\eta$}
	\label{algo:gradient-descent}
	\begin{algorithmic}\small
	    \State \textbf{Input:} Obtain $(\vu_0,\vv_0)$ via Algorithm \ref{algo:initialization} below.
		\For{$t = 1, \ldots$}
	    \State $ \vu_t \gets  \vu_{t-1} - \eta \nabla \tf_{\vh} (\vu_{t-1},\vv_{t-1} )$
		\State $ \vv_t \gets  \vv_{t-1} - \eta \nabla \tf_{\vm} (\vu_{t-1},\vv_{t-1} )$
		\EndFor
	\end{algorithmic}
\end{algorithm}

Finally, a suitable initialization $(\vu_0,\vv_0)$ for Algorithm \ref{algo:gradient-descent} is computed using Algorithm \ref{algo:initialization} below. In short, the left and right singular vectors of $\setA^*(\hat{\vy})$ when projected in the set of sufficiently incoherent (measured in terms of the coherence $\mu$, $\nu$ of the original vectors $\vh_0$, and $\vx_0$) vectors supply us with the initializers $(\vu_0, \vv_0)$. 
\begin{algorithm}[H]
	\caption{Initialization}\small
	\label{algo:initialization}
	\begin{algorithmic}
		\State \textbf{Input:} Compute $\setA^*(\hat{\vy})$, and find the leading singular value $d$, and the corresponding left and right singular vectors $\hat{\vh}_0$, and $\hat{\vx}_0$, respectively. 
		\State Solve the following optimization programs 
		\State $\vu_0 \gets \underset{\vh}{\text{argmin}} \ \|\vh - \sqrt{d} \hat{\vh}_0\|_2, \ \text{subject to} \  \sqrt{Q}\|\mF_M\vh\|_\infty \leq 2 \sqrt{d} \mu,$\ \text{and} \ 
		\State $\vv_0 \gets \underset{\vx}{\text{argmin}} \ \|\vx - \sqrt{d} \hat{\vx}_0\|_2, \ \text{subject to} \ \sqrt{Q}\|\mF_K^*\vx\|_\infty \leq 2 \sqrt{d} \nu.$
		\State \textbf{Output:} $(\vu_0,\vv_0)$. 
	\end{algorithmic}
\end{algorithm}

\subsection{Main Results} 
Our main result shows that given the convolution measurements \eqref{eq:measurements}, a \textit{suitably initialized} Wirtinger gradient-descent Algorithm \ref{algo:gradient-descent} converges to the true solution, i.e., $(\vu_t, \vv_t) \approx (\vh_0, \vx_0)$ under an appropriate choice of $Q$. To state the main theorem, we need to introduce some neighborhood sets. For vectors $\vh \in \C^M$, and $\vx \in \C^{K}$, we define the following sets of neighboring points of $(\vh,\vx)$ based on either, magnitude, coherence, or the distance from the ground truth.
\begin{align}
&\setN_{d_0} := \{(\vh,\vx) | \|\vh\|_2 \leq 2\sqrt{d_0}, \ \|\vx\|_2\leq 2 \sqrt{d_0} \},\label{eq:setNd-def}\\
&\setN_\mu := \{ (\vh,\vx) | \sqrt{Q}\|\mF_M\vh\|_\infty \leq 4\mu\sqrt{d_0} \},\label{eq:setNmu-def}\\
&\setN_\nu : = \{(\vh,\vx)|  \sqrt{Q}\|\mF_K^*\vx\|_\infty \leq 4\nu\sqrt{d_0} \},\label{eq:setNnu-def}\\
&\setN_{\varepsilon} := \{(\vh,\vx) | \|\vh\vx^*-\vh_0\vx_0^*\|_{F} \leq \varepsilon d_0 \}\label{eq:setNe-def}.
\end{align}

Our main result on blind deconvolution from modulated inputs \eqref{eq:measurements} is stated below.  
\begin{thm}[Theorem 1 in \cite{Ahmed2018Modulation} ]\label{thm:convergence}
	 Fix $0 < \varepsilon \leq 1/15$. Set $\vs_{0} = \mF_K^*\vx_{0} \in \C^Q$; and $\vx_{0} \in \C^K$, and $\vh_0 \in \C^M$ be arbitrary vectors. Assume $Q \geq M$. Let the signal coherence parameters be as defined in \eqref{eq:mu-nu}.  Let $\vr$ be independently generated $Q$-vector with standard iid Rademacher entries. We observe the $Q$-point circular convolutions of the random sign vectors $\vr\odot\vs_{0}$ with $\vh_0$, leading to observations \eqref{eq:measurements} contaminated with additive noise $\ve$. Assume that the initial guess $(\vu_0,\vv_0)$ of $(\vh_0, \vx_0)$ belongs to $\tfrac{1}{\sqrt{3}}\setN_{d_0} \cap \tfrac{1}{\sqrt{3}}\setN_{\mu}\cap \tfrac{1}{\sqrt{3}}\setN_{\nu} \cap \setN_{\frac{2}{5}\varepsilon},$ and that
	 \begin{align}\label{eq:sample-complexity-main-thm}
	 Q \geq c\frac{\nu^2}{\varepsilon^4} (\mu^2 K + \nu^2 M)\log^4(Q),
	 \end{align}
	then Algorithm \ref{algo:gradient-descent} will create a sequence $(\vu_t,\vv_t) \in \setN_{d_0}\cap \setN_{\mu} \cap \setN_{\nu} \cap \setN_{\varepsilon}$, which converges geometrically to $(\vh_0,\vx_0)$ with probability at least 
	\begin{align}\label{eq:probability-main-thm}
	1-2\exp\left(-c\varepsilon^4Q/\mu^2\nu^4\right),
	\end{align}
	and there holds 
	\begin{align}\label{eq:stable-recovery-bound}
	&\max\{ \sin \angle (\vu_t,\vh_0), \sin \angle (\vv_t, \vx_0)\} \leq \notag\\
	&\qquad\qquad \frac{1}{d_t} \Big(\tfrac{2}{3}(1-\eta\omega)^{t/2}\varepsilon d_0
	+ 50 \|\setA^*(\ve)\|_{2\rightarrow 2}\Big),
	\end{align}
	 and $
	 |d_t-d_0| \leq \tfrac{2}{3} (1-\eta \omega)^{t/2} \varepsilon d_0 + 50 \|\setA^*(\ve)\|_{2\rightarrow 2},$
	where $d_t = \|\vu_t\|_2\|\vv_t\|_2$, $\omega > 0$, $\eta$ is the fixed step size. Fix $\alpha \geq 1$. For noise $\ve \sim \text{Normal}(\mathbf{0},\frac{\sigma^2 d_0^2}{2Q}\mI_{Q}) + \iota \text{Normal}(\mathbf{0},\frac{\sigma^2 d_0^2}{2Q}\mI_{Q})$, $\|\setA^*(\ve)\|_{2\rightarrow 2} \leq \frac{2\varepsilon}{50} d_0$ with probability at least $1-\setO((Q)^{-\alpha})$ whenever 
	\begin{align}\label{eq:sample-complexity-LN}
	Q \geq  c_\alpha\frac{\sigma^2}{\varepsilon^2}  \max(M,K\log(Q))\log(Q).
	\end{align}
\end{thm}
Proof of the theorem above can be found in the preprint available on arxiv \cite{Ahmed2018Modulation}. 
The above theorem claims that starting from a good enough initial guess the gradient descent algorithm converges super linearly to the ground truth. The theorem below guarantees that the required good enough initialization: $(\vu_0,\vv_0) \in \frac{1}{\sqrt{3}}\setN_{d_0} \cap \frac{1}{\sqrt{3}}\setN_{\mu}  \cap \frac{1}{\sqrt{3}}\setN_{\nu}\cap \setN_{\frac{2}{5}\varepsilon}$ is supplied by Algorithm \ref{algo:initialization}. 
\begin{thm}[Theorem 2 in \cite{Ahmed2018Modulation}]\label{thm:initialization}
	The initialization obtained via Algorithm \ref{algo:initialization} satisfies $
	(\vu_0,\vv_0) \in\frac{1}{\sqrt{3}}\setN_{d_0} \cap \frac{1}{\sqrt{3}}\setN_{\mu}  \cap \frac{1}{\sqrt{3}}\setN_{\nu}\cap \setN_{\frac{2}{5}\varepsilon},$ and $0.9 d_0 \leq d \leq 1.1 d_0$ 	holds with probability at least $1-2\exp\left(-c\varepsilon^2 Q/\mu^2\nu^4\right)$ whenever 
	\[
	Q \geq c \frac{\nu^2}{\varepsilon^2} \left(\mu^2\nu_{\max}^2 K + \nu^2 M\right) \log^4 Q.
	\]
\end{thm}
Proof of the theorem above can be found in the preprint available on arxiv \cite{Ahmed2018Modulation}. 
\noindent

\subsection{Discussion}\label{sec:discussion}

Theorem \ref{thm:convergence}, and \ref{thm:initialization} together prove that randomly modulated unknown  $Q$-vector $\vs_{0}$, and unknown $M$-vector $\vh_0$ can be recovered from their circular convolutions $\vh_0 \circledast (\vr\odot\vs_{0}),$ under suitably large $Q$. We will refer to the bounds in \eqref{eq:sample-complexity-main-thm}, \eqref{eq:sample-complexity-LN} as \textit{sample complexity bounds}. 
Observe that the number of unknowns in the system of equations \eqref{eq:measurements} is $K+M$. Combining \eqref{eq:sample-complexity-main-thm}, and \eqref{eq:sample-complexity-LN}, it becomes clear that number $Q$ of measurements required for successful recovery scale with $K+M$ (within coherences, and log factors). The bound on $Q$ above is information theoretically optimal (within log factors and coherence terms). This sample complexity result almost matches the results in \cite{ahmed2012blind,li2016rapid} except for an extra $\nu^2$, and a log factor. However, the important difference is, as mentioned in the introduction, that unlike  \cite{ahmed2012blind,ahmed2016leveraging,li2016rapid,ahmed2018convex,lee2017spectral}, the inputs are not assumed to reside in Gaussian subspaces rather only have random signs. 

A direct application of our main result shows that $\vs$, and hence $s(t)$ can be recovered from the received signal $y(t) = (s(t)\odot r(t)) \circledast h(t)$ without knowing channel impulse response by operating the random binary waveform $r(t)$ at rate $Q \gtrsim \nu^2(\mu^2K+\nu^2M)\log^4 Q$, and sampling the received signal $y(t)$ at a rate $Q$. The coherences $\nu^2$, and $\mu^2$ are simply the peak values in time $ \|\vs\|_\infty^2$, and frequency domain $\|\mF_M\vh\|_\infty^2$, respectively.

\section{Numerical Simulations}\label{sec:numerics}

In this section, we numerically investigate the sample complexity bounds using phase transitions. We also report stable recovery in the presence of additive measurement noise. 

 We present phase transitions to numerically investigate the constraints in \eqref{eq:sample-complexity-main-thm}, and \eqref{eq:sample-complexity-LN} on the dimensions $Q$, $M$, and $K$ for the gradient descent algorithm to succeed with high probability. The shade represents the probability of failure, which is computed over hundred independent experiments. For each experiment, we generate Gaussian random vectors $\vh_0$, and $\vx_{0}$, and generate $\vs_{0} = \mF_K^*\vx_0$. The synthetic measurements are then generated following the model \eqref{eq:model}. We run Algorithm \ref{algo:gradient-descent} initialized via Algorithm \ref{algo:initialization}, and classify the experiment as successful if the relative error 
\begin{align}\label{eq:relative-error}
\text{Relative Error}: = \frac{\|\hat{\vh}\hat{\vx}^*-\vh_0\vx_0^*\|_F}{\|\vh_0\vx_0^*\|_F}
\end{align}
is below $10^{-2}.$ The probability of success at each point is computed over hundred such independent experiments. 

Phase diagram in Figure \ref{fig:Numerics}(a) investigate successful recovery for a fixed $Q = 3200$, and under varying $K$, and $M$. The shade in the phase transition represents probability of failure. For example, in the phase transition, successful recovery occurs almost always when the measurements are a factor of $2.8$ above the number of unknowns, that is, $Q \geq 2.8(K+M)$.

 Noise performance of the algorithm is depicted in Figure \ref{fig:Numerics}(b). Additive Gaussian noise $\ve$ is added in the measurements as in \eqref{eq:measurements}. As before, we synthetically  generate $\vh_0$, and $\vx_0$ as Gaussian vectors. We plot (second) relative error (log scale) in \eqref{eq:relative-error} of the recovered vectors $\hat{\vh}$, and $\hat{\vx}$ averaged over hundred independent experiments vs. $
\text{SNR} := 10 \log_{10} \left(\|\vh_0\vx_0^*\|_F^2/\|\ve\|_2^2\right)$
, and Figure \ref{fig:Numerics}(c) shows average relative error (log scale) vs. oversampling ratio := $Q /(K+M)$ under no noise. Oversampling ratio is a factor by which the number ($Q$  measurements exceed the number $K+M$ of unknowns. The second plot shows that the relative error degrades \textit{gracefully} by reducing SNR, and the third shows that relative error almost reduces to zero when the oversampling ratio exceeds 2.1. 
\bibliographystyle{IEEEbib}
\bibliography{IEEEabrv,CPref}

\end{document}